\documentclass{aastex}
\usepackage{amssymb}
\usepackage{epsfig}
\usepackage{graphicx}
\usepackage{longtable}
\usepackage{caption}
\usepackage{amsfonts}
\usepackage{amscd}
\usepackage{graphics}
\usepackage{amsmath}
\usepackage{times}
\usepackage{color}
\usepackage{footnote}
\usepackage{ulem}
\usepackage{subfig}
\usepackage[T1]{fontenc}
\usepackage{enumerate}
\usepackage{multicol}
\usepackage{multirow}
\shorttitle{Broadband spectral study of VHE blazar 1ES\,1011+496}
\shortauthors{Sinha et. al}

\begin{document}

\title{ On the spectral curvature of  VHE blazar 1ES\,1011+496: Effect of spatial particle diffusion}
\author{Atreyee Sinha\altaffilmark{1}, S. Sahayanathan\altaffilmark{2}, B. S. Acharya\altaffilmark{1}, G. C. Anupama\altaffilmark{3},V. R. Chitnis\altaffilmark{1}, B. B. Singh\altaffilmark{1}}

\altaffiltext{1}{Department of High Energy Physics, Tata Institute of Fundamental Research, Homi Bhabha Road, 
	Mumbai 400005, INDIA; atreyee@tifr.res.in}
\altaffiltext{2}{Nuclear Research Laboratory, Bhabha Atomic Research Center, Mumbai, India}
\altaffiltext{3}{Indian Institute of Astrophysics, II Block, Koramangala, Bangalore, 560 034, India}

\begin{abstract}

	A detailed multi-epoch study of the broadband spectral behaviour of the very high energy (VHE) source, 1ES\,1011+496, provides us with 
	valuable information regarding the underlying particle distribution. Simultaneous observations of the source at optical/ UV/ X-ray/ $\gamma$-ray during three different epochs, as obtained from Swift-UVOT/ Swift-XRT/ Fermi-LAT, are supplemented with the information available from the VHE telescope array, HAGAR. The longterm flux variability at the Fermi-LAT energies is clearly found to be lognormal. It is seen that the broadband spectral energy distribution (SED) of 1ES\,1011+496 can be successfully reproduced by synchrotron and synchrotron self Compton emission models. Notably, the observed curvature in the photon spectrum at X-ray energies demands a smooth transition of the 
underlying particle distribution from a simple power law to a power law with an exponential cutoff or a smooth broken 
power law distribution, which may possibly arise when the escape of the particles from the main
	emission region is energy dependent. Specifically, if the particle escape rate is related to its energy as $E^{0.5}$ then the observed photon 
spectrum is consistent with the ones observed during the various epochs.

\end{abstract}
\keywords{BL Lacertae objects: individual (1ES1011+496)- galaxies: active - X-rays: galaxies - radiation mechanisms: non-thermal}

\maketitle


\section[sect:Intro]{Introduction}
Blazars are a peculiar subclass of radio loud active galactic nuclei (AGN) where a powerful relativistic jet is pointed close to the line of sight of the observer \citep{UrryPadovani}. They show high optical polarization, intense and highly variable non-thermal radiation throughout
the entire electromagnetic spectra in time scales extending from minutes to years, apparent super-luminal motion in
high resolution radio maps, large Doppler factors and beaming effects. Blazars can be broadly classified into two sub-groups, BL~Lacs and flat spectrum radio quasars (FSRQs), where the former are identified by the absence of  emission/absorption lines. The broadband spectral energy distribution (SED) of blazars is characterized by two peaks, one in the IR - X-ray regime, and the second one in $\gamma$-ray regime. According to the location of the first peak, BL~Lacs are further classified into Low energy peaked BL~Lacs (LBLs), Intermediate energy peaked BL~Lacs (IBLs) and High energy peaked BL~Lacs (HBLs) \citep{PadGio}. Both leptonic (eg: \cite{leptonic_maraschi,leptonic_dermer,leptonic_sikora,leptonic_bloom,leptonic_bla}) and hadronic (eg: \cite{had1,had2,had3}) models have been proposed to explain the broadband SED with
varying degrees of success. While the origin of the low energy component is well established to be caused by synchrotron emission from relativistic electrons gyrating in the magnetic field of the jet, the physical mechanisms responsible for the high energy emission are still under debate. It can be produced either via inverse Compton scattering (IC) of low frequency photons by the same
electrons responsible for the synchrotron emission (leptonic models), or via hadronic processes initiated by relativistic
protons, neutral and charged pion decays or muon cascades (hadronic models). The seed photons for IC in leptonic
models can be either the source synchrotron photons (Synchrotron Self Compton, SSC) or from external sources such
as the Broad Line Region (BLR), the accretion disc, the cosmic microwave background, etc (External Compton, EC).  For a comprehensive review of these mechanisms, see \cite{bottcher2007}. 

1ES\,1011+496 (RA = 10:15:04.14, Dec = 49:26:00.70; J2000) is a HBL located at a redshift of $z=0.212$. It was discovered as a VHE emitter by the MAGIC collaboration in 2007, following an optical outburst in March 2007 \citep{1es1011_Magic}. The flux above 200 GeV was roughly $7\%$ of the Crab Nebula, and the observed spectrum was reported to be a power law with a very steep index of $4.0 \pm 0.5$. After correction for attenuation of VHE photons by the extragalactic background light (EBL; \cite{kneiske}), the intrinsic spectral index was computed to be $3.3 \pm 0.7$. At its epoch of discovery, it was the most distant TeV source. \cite{1es1011_Magic} had constructed the SED with simultaneous optical R-band data, and other historical data from \cite{costamante} and modelled it with a single zone radiating via SSC processes. However, the model parameters could not be constrained due to the sparse sampling and the non-simultaneity of the data. 
\cite{egret3C} had suggested the association of this source with the EGRET source 3EG J1009+4855, but this  association has later been challenged \citep{egret_NS}. This source has been detected in the $0.1 - 300$ GeV  band by Fermi-LAT, and in the $0.3-10$ keV band by Swift-XRT \citep{Abdo_SED}. A detailed study of its optical spectral variability has been done by \cite{bottcher_1011}. Results of multiwavelength campaigns carried in 2008 \citep{1011_2008} and 2011-2012 \citep{1011_2011-12} have recently been published  by the MAGIC collaboration.

In February 2014, 1ES\,1011+496 was reported to be in its highest flux state till date as seen by Fermi-LAT \citep{ATEL5888}, and Swift-XRT \citep{ATEL5866}. During this time, the VERITAS collaboration also detected a strong VHE flare from this source, at an integral flux level of $\sim 20\% \ {\rm{to}} \ 75\% $ of the Crab flux, which was almost a factor of 10 higher than its baseline flux \citep{VERITAS_update}. \cite{1011_2014} have used the TeV spectrum during this flare to put contrains on the EBL density. This source was observed by the High Altitude GAmma Ray (HAGAR) Telescope array during the February-March 2014 season. In this paper, we study the simultaneous SED of this source as seen by Swift, Fermi-LAT and HAGAR during this epoch. To understand the broadband spectral behaviour, we also construct SEDs using quasi-simultaneous data of two previous epochs.

In Section \ref{sect:Analysis}, we describe our data reduction procedure and study the temporal variability in the lightcurves. The distribution function of the flux at the $\gamma$-ray energies is studied in Section \ref{sect:lognorm}. We model the SED using a SSC model having a smoothly varying power law spectrum of the underlying electron energy distribution, and the results are outlined in Section \ref{sect:SED}. We discuss the implications in Section \ref{sect:disc}, and show that such a situation may arise when the escape of the particles from the emission region is energy dependent. The results are summarized in Section \ref{sect:conclusion}. A cosmology with $\omega_m = 0.3$, $\omega_{\Lambda} = 0.7$ and $H_{0} = 70$ $km$ $s^{-1}$ $Mpc$ is used in this work.

\section[sect:Analysis]{Data analysis and lightcurves}\label{sect:Analysis}

\subsection[sect:fermi]{Fermi-LAT}\label{sect:fermi}

Fermi-LAT data are extracted from a region of 20$^{\circ}$ centered on the source. The standard data analysis procedure as mentioned in the Fermi-LAT  documentation\footnote{\url{http://fermi.gsfc.nasa.gov/ssc/data/analysis/documentation/}}
is used. Events belonging to the energy range 0.2$-$300 GeV and SOURCE class are used. To select good time intervals, 
a filter ``\texttt{DATA$\_$QUAL$>$0}'', \&\& ``\texttt{LAT$\_$CONFIG==1}'' is used and only events with less than 105$^{\circ}$ zenith angle are selected to avoid contamination from the Earth limb $\gamma$-rays. 
The galactic diffuse emission component gll\_iem\_v05\_rev1.fits and an isotropic component iso\_source\_v05\_rev1.txt are used as the background models. The unbinned likelihood method included in the pylikelihood library of {\tt Science Tools (v9r33p0)} and the post-launch instrument response functions P7REP\_SOURCE\_V15 are used for the analysis. All the sources lying within 10$^{\circ}$ region of interest (ROI) centered at the position of 1ES\,1011$+$496 and 
defined in the third {\it Fermi}-LAT catalog \citep{3FGL}, are included in the xml file. All the parameters except the scaling factor of the sources within the ROI are allowed to vary during the likelihood fitting. The source spectrum is assumed to be a power law.

\begin{figure*}
	\centering
	\includegraphics[scale=0.8]{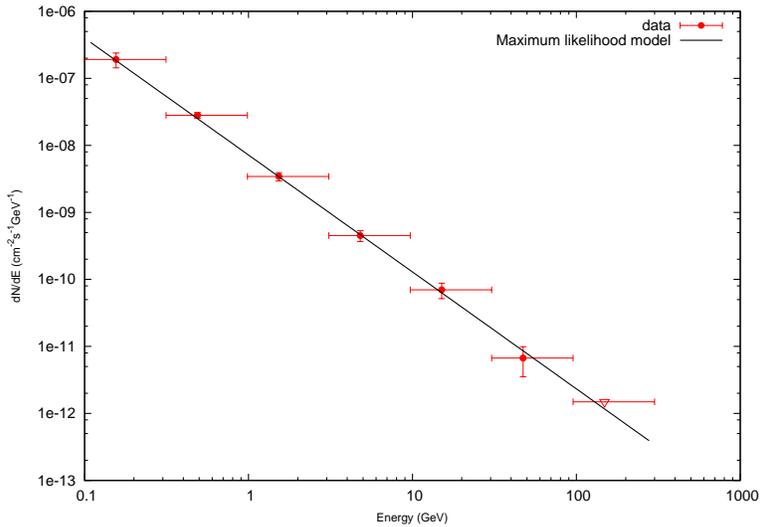}
	\caption{Energy spectrum of 1ES1\,1011+496 from six years of Fermi-LAT data during 2008-2014. The last point is an upper limit and is shown by an inverted triangle. The spectrum is well fit by a power law of index $\alpha = 1.82 \pm 0.01$.}
	\label{fermi_spec}
\end{figure*}

Analysis of all data for this source from 2008 - 2014 yields a spectrum consistent with a simple power law with an index of $1.82 \pm 0.01 $ (Figure \ref{fermi_spec}), and the flux is found to be variable on a time scale of ten days with a significance of $12.7 \sigma$. The fractional variability amplitude parameter \citep{Vaughan,varsha_var} is computed to be $F_{var}=0.35 \pm 0.01$, where 

\begin{equation} 
        F_{var}=\sqrt{\frac{S^2-\sigma^2_{err}}{\bar{x}^2}}
 \end{equation} 
\noindent
Here $\sigma^2_{err}$ is the mean square error, $\bar{x}$ the unweighted sample mean, and $S^2$ the sample variance, and the error on $F_{var}$ is given as 
\begin{equation}
        \sigma_{F_{var}}= \sqrt{ \left( \sqrt{\frac{1}{2N}}\cdot\frac{\sigma^2_{err}}{\bar{x}^2 F_{var}} \right)^2 +  \left( \sqrt{\frac{\sigma^2_{err}}{N}}\cdot\frac{1}{\bar{x}} \right)^2}
.\end{equation}
\noindent
with $N$ as the number of points. 

\begin{figure*}
	\centering
	\includegraphics[scale=1.3]{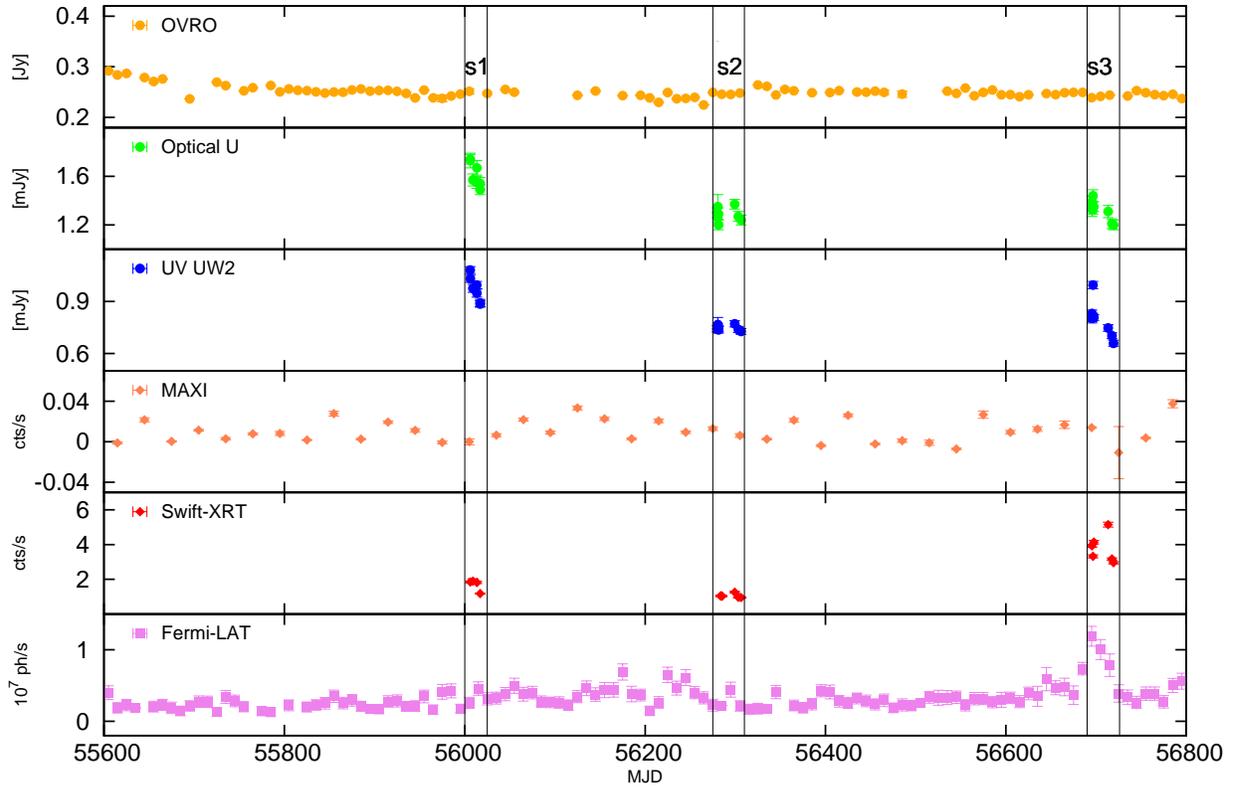}
	\caption{Multiwavelength lightcurve during MJD 55600 to 56800 (calendar days 2011-2014) showing from top:
	Panel 1: OVRO flux in Jy; Panel 2: Optical U band flux in mJy; Panel 3: UV flux in UW2 band in mJy;
	Panel 4: MAXI flux (with monthly binning) in counts/sec; Panel 5: {\it Swift}-XRT flux in counts/sec; 
Panel 6: {\it Fermi}-LAT flux (with 10 days binning) in $ph/cm^2/sec$. The three states for which SED has been studied are marked and labelled as s1, s2 and s3 respectively.}
	\label{fig:lc}
\end{figure*}

  There is no significant trend of spectral hardening with increasing flux (Spearman's rank correlation, $rs=-0.25$), which has been seen in many HBL. The light curve for 3 years period during 2011 to 2014 is shown in the bottom panel of Fig \ref{fig:lc}. 

  Spectra are extracted in five logarithmically binned energy bins for three epochs contemporaneous with Swift observations, corresponding to MJD (a) 56005 to 56020 (state s1) (b) 56280 to 56310 (state s2) and (c) 56692 to  56720 (state s3). LAT fluxes and spectral parameters during these epochs are given in Table \ref{tab:states}. The state s3 corresponds to the period for which the highest gamma ray flux from this source is seen till date.

\subsection{Swift-XRT}

\begin{figure*}
	\centering
	\includegraphics[scale=0.4,angle=180]{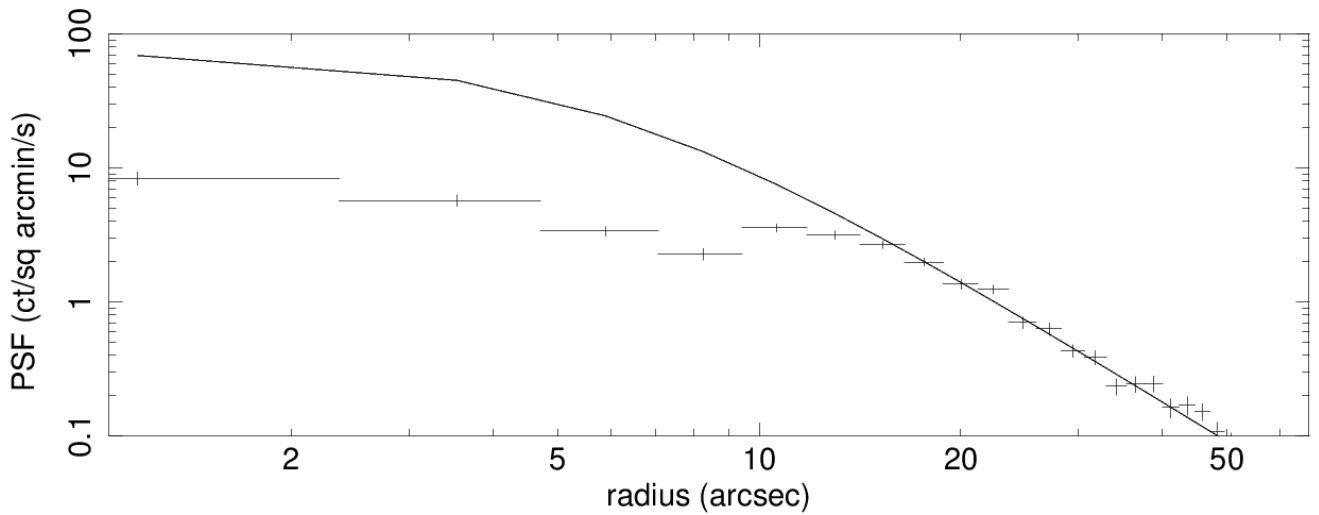}
	\caption{XRT PSF for obs 00035012032 fitted by a Kings function. The deviation from the model is seen for regions smaller than 16 arc seconds, which are thus excluded from the source region.}
	\label{fig:psf}
\end{figure*}

 A total of 16 Swift pointings are available during the studied epochs, the ids of which are given in Table \ref{tab:states}. Swift-XRT data \citep{XRT_instru} are processed with the XRTDAS software package (v.3.0.0) available within HEASOFT package (6.16). Event files are cleaned and calibrated using standard procedures ({\tt xrtpipeline v.0.13.0}), and {\tt xrtproducts v.0.4.2} is used to obtain the lightcurves and spectra. Observations are available both in Windowed Timing (WT) and Photon Counting (PC) modes, and full grade selections (0-2 for WT and 0-12 for PC) are used. PC observations during the 2014 flare are heavily piled up (counts $> 0.5$ c/s), and  is corrected for by following the procedure outlined in the Swift analysis threads\footnote{\url{http://www.swift.ac.uk/analysis/xrt/pileup.php}}. The XRT Point Spread Function is modelled by a King function 
\begin{equation}
	PSF(r) = [1 + (r/r_c)^2]^{-\beta}
\end{equation}
with $r_c = 5.8 $ and $\beta = 1.55 $ \citep{XRTPSF}. Depending on the source brightness, annular regions are chosen to exclude pixels deviating from the King's function. The tool {\tt xrtmkarf} is then executed  with PSF correction set to ``yes" to create an ARF corrected for the loss of counts due to the exclusion of this central region. For eg, for observation id 00035012032 (see Figure \ref{fig:psf}), an annular region of 16-25 arc seconds centered on the source position is taken as the source region.

The lightcurves are finally corrected for telescope vignetting and PSF losses with the tool {\tt xrtlccorr v.0.3.8}. The spectra are combined using the tool {\tt addspec} for all observations within each of the states s1, s2 and s3 as defined in \ref{sect:fermi}. Spectra are grouped to ensure a minimum of 30 counts in each bin by using the tool {\tt grppha v.3.0.1}. 

A slight curvature is detected in the XRT spectrum, and a log parabolic spectral model given by 
\begin{equation}
         dN/dE = K(E/E_b)^{-\alpha -\beta log(E/E_{b})}
 ,\end{equation}
\noindent
is used to model the observed spectrum. Here, $\alpha$ gives the spectral index at $E_{b}$, which is fixed at $1keV$ during the fitting. Parameters obtained during the fitting are given in Table \ref{tab:states}. To correct for the line of sight absorption of soft X-rays due to the interstellar gas, the neutral hydrogen column density is fixed at $N_{H}= 8.38 \times 10^{19} cm^{-2} $ \citep{LAB}.

\subsection{Swift-UVOT}

Swift-UVOT observations cycled through the 6 filters, the optical U, V, B, and the UV UW1, UW2 and UM2. The individual exposures during each of the states are summed using {\tt uvotimsum v.1.6}, and {\tt uvotsource v.3.3} tool is used to extract the fluxes from the images using aperture photometry. The observed fluxes are corrected for galactic extinction using the dust maps of \cite{schelgel}, and for contribution from the host galaxy following \cite{host_corr}, with a R-mag $ = 16.41 \pm 0.09$.  

\subsection{Other multiwavelength data}

\begin{figure*}
	\centering
	\subfloat[Cutoff power law]{\includegraphics[scale=0.55]{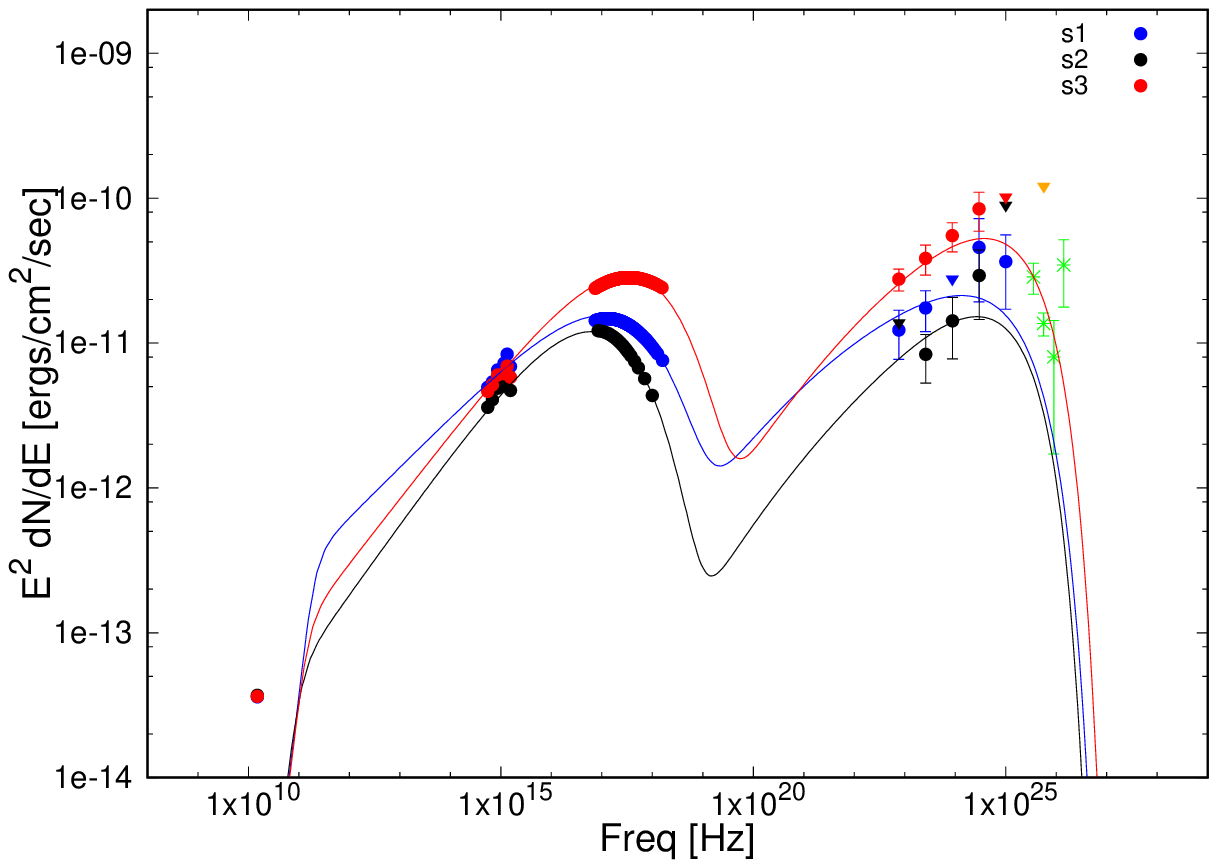}}
	\qquad
	\subfloat[Smooth broken power law]{\includegraphics[scale=0.55]{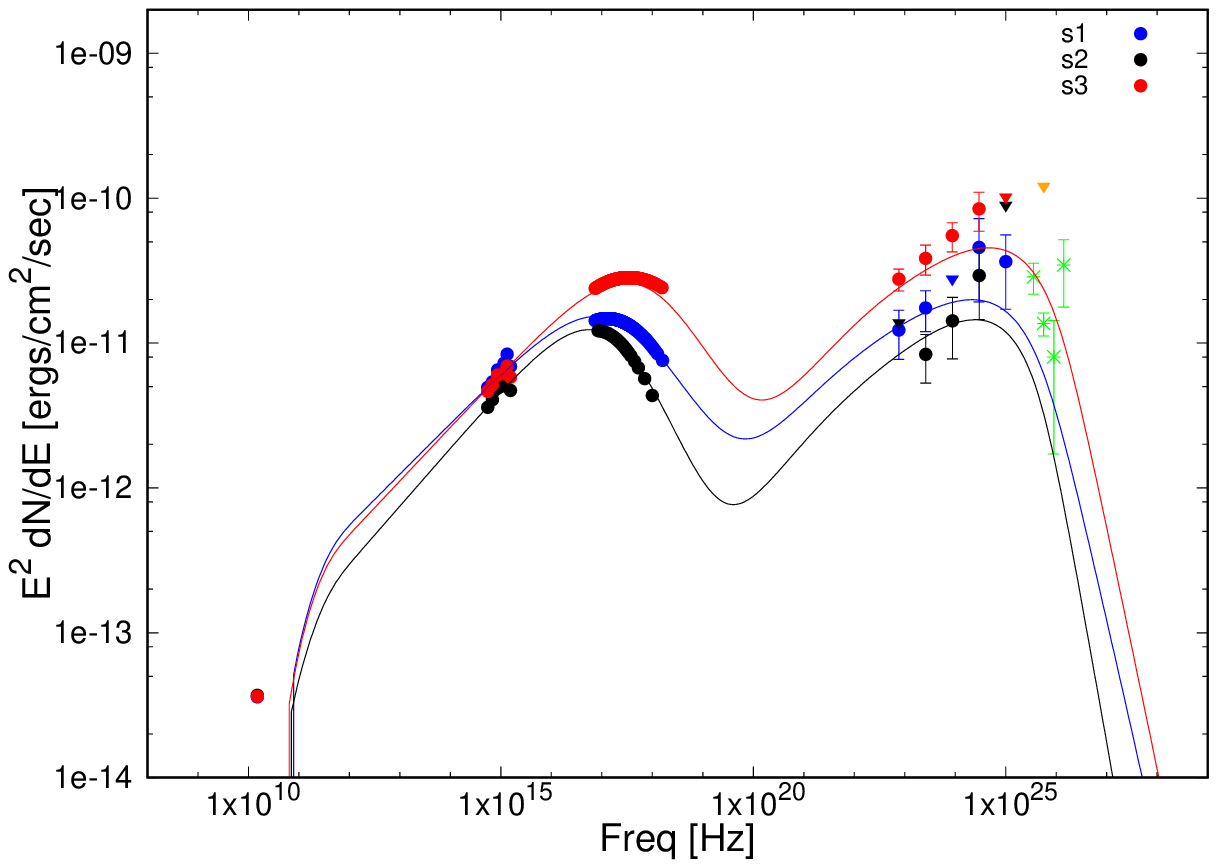}}
	\caption{Spectral energy distribution of 1ES\,1011+496 during the 3 epochs studied in the paper, with simultaneous data from Swift-UVOT, Swift-XRT and the Fermi-LAT. The orange inverted triangle gives the HAGAR upper limit during the Feb-March, 2014 season. The green stars show the MAGIC spectrum during its discovery in 2007 (\citet{1es1011_Magic}). The SEDs are modelled with a one zone SSC with the underlying electron distribution as (a) A power law with exponential cutoff and (b) A smooth broken power law.}
	\label{SED}
\end{figure*}

We supplement the above information with other multiwavelength flux measurements at different energies:

\begin{enumerate}[(i)]
	
	\item {\bf Radio} \hfill \\
		As a part of the Fermi monitoring program, the Owens Valley Radio Observatory (OVRO) \cite{ovro} has been regularly observing this source since 2008. Flux measurements at 15 GHz taken directly from their website\footnote{\url{http://www.astro.caltech.edu/ovroblazars/}} show negligible variability with $f_{var} \sim 0.07$.

	\item {\bf X-ray} \hfill \\
		Daily binned source counts in the $2-20$ keV range from the  Monitor of All-sky X-ray Image (MAXI) on board the International Space Station (ISS; \cite{MAXI})  are available from their website\footnote{\url{http://maxi.riken.jp/}}. The X-ray counts binned on monthly timescales show high variability, with $f_{var}=1.34 \pm 0.08$.

	\item {\bf VHE} \hfill\\
		HAGAR \citep{kiranHAGAR} is a hexagonal array of seven Atmospheric Cherenkov Telescopes (ACT) which uses the wavefront sampling technique to detect celestial  gamma rays. It is located at the Indian Astronomical Observatory site (32$^{\circ}$46'46" N, 78$^{\circ}$58'35" E), in Hanle, Ladakh in the Himalayan mountain ranges at an altitude of 4270m. The energy threshold for the HAGAR array for vertically incident $\gamma$-ray showers is 208 GeV, with a sensitivity of detecting a Crab Nebula like source in 17hr for a 5$\sigma$ significance. Detailed descriptions of HAGAR instrumentation and simulations can be found in  \cite{Amit421} and  \cite{labHAGAR}. HAGAR observations of 1ES\,1011+496 were carried out during the February-March 2014 season following an alert by MAGIC and VERITAS collaborations of a VHE flare from this source during 3 February to 11 February \citep{ATEL5887}. The observations were carried out in clear moon-less conditions, between 19 February to 8 March, 2014 (MJD 56707 to 56724). Each pointing source (ON) run of approximately 60 mins duration was followed (or preceded) by a background (OFF) run of the the same time duration at the same zenith angle and having similar night sky brightness as source region. A total of 23 run  pairs are taken corresponding to a total duration of 1035 mins with common ON-OFF hour angle.		
		
		The data are reduced following the procedures outlined in \cite{Amit421}, with various quality cuts imposed. Only events with signals seen in at least 5 telescopes are retained, for which the energy threshold is calculated to be 234 GeV. No significant signal is seen from the source with 600 mins of clean data. The excess of signal over background is computed to be $706 \pm 566$ photons, corresponding to a significance of $1.3 \sigma$. A $3-\sigma$ upper limit of $1.2 * 10^{-10} ergs/cm^2/ sec$ for the flux of gamma rays above 234 GeV is calculated from the above data, which corresponds to roughly $70\%$ of the Crab Nebula flux. 

		This is supplemented with VHE spectra of this source obtained during its epoch of discovery \citep{1es1011_Magic} by the Major Atmospheric Gamma-ray Imaging Cherenkov (MAGIC).  These spectral points have been plotted for representative purposes in Figure \ref{SED} and provide a lower limit for the SED modelling during the VHE flare in 2014, epoch s3.

\end{enumerate}

\section[sect:Lognorm]{Detection of lognormality}\label{sect:lognorm}

 Log-normal flux distributions and linear rms-flux relations have often been claimed to be an universal feature of accretion powered sources like X-ray binaries \citep{lognorm_xrb,scaringi}. Lognormal fluxes have fluctuations, that are, on average, proportional to the flux itself, and are indicative of an underlying multiplicative, rather than additive physical process. It has been suggested that a lognormal flux behaviour in blazars could be indicative of the variability imprint of the accretion disk onto the jet \citep{blazar_var}. This behaviour in blazars was first clearly detected in the X-ray regime in BL~Lac \citep{giebels_lognorm} and has hence been seen across the entire electromagnetic spectrum in PKS~2155$-$304 \citep{lognorm_2155} and Mkn~421 \citep{421_longterm}.

\begin{figure*}
	\centering
	\subfloat[Flux distribution]{\includegraphics[scale=0.35]{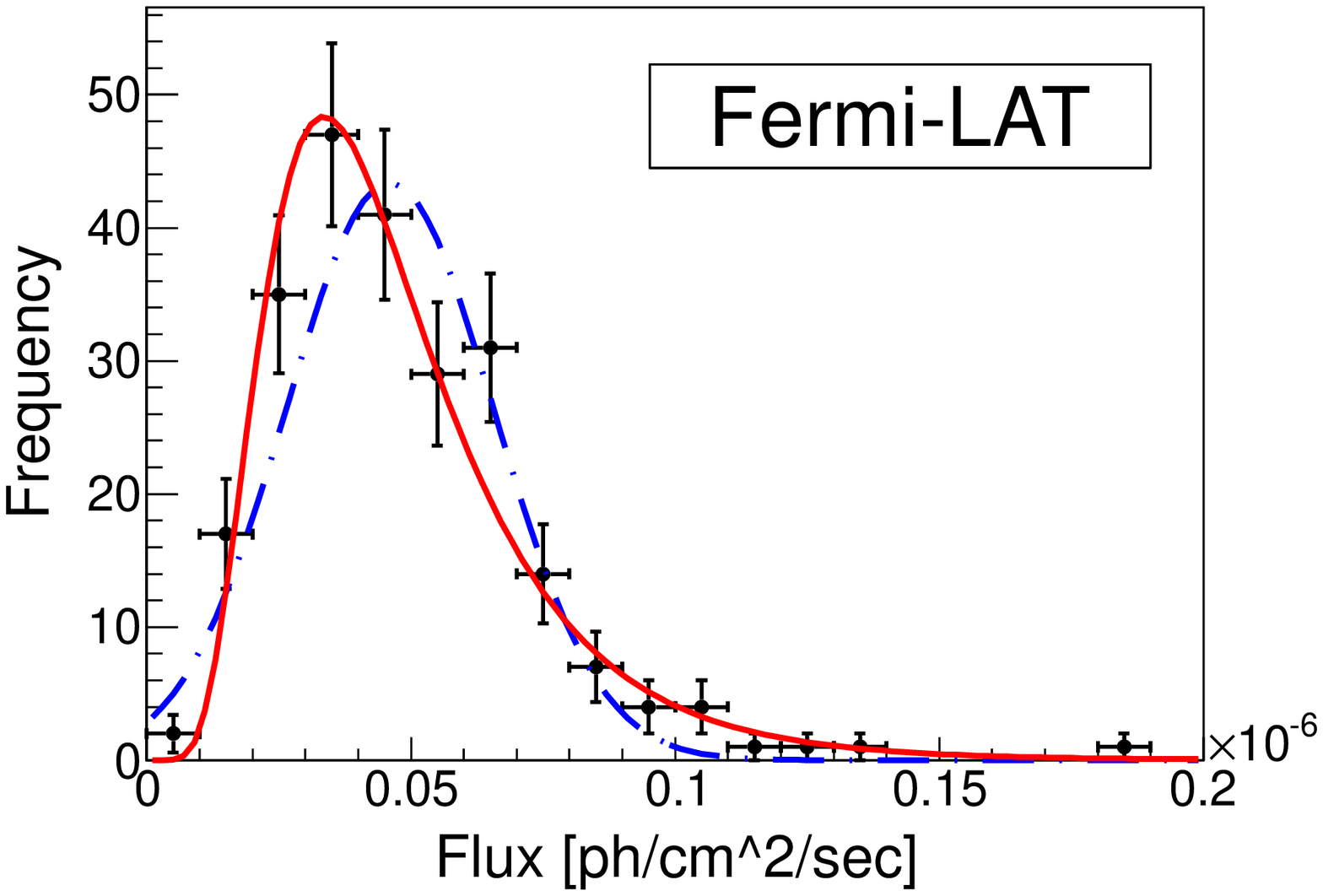}\label{Fig:hist}}
	\qquad
	\subfloat[Flux-rms relation]{\includegraphics[scale=0.55]{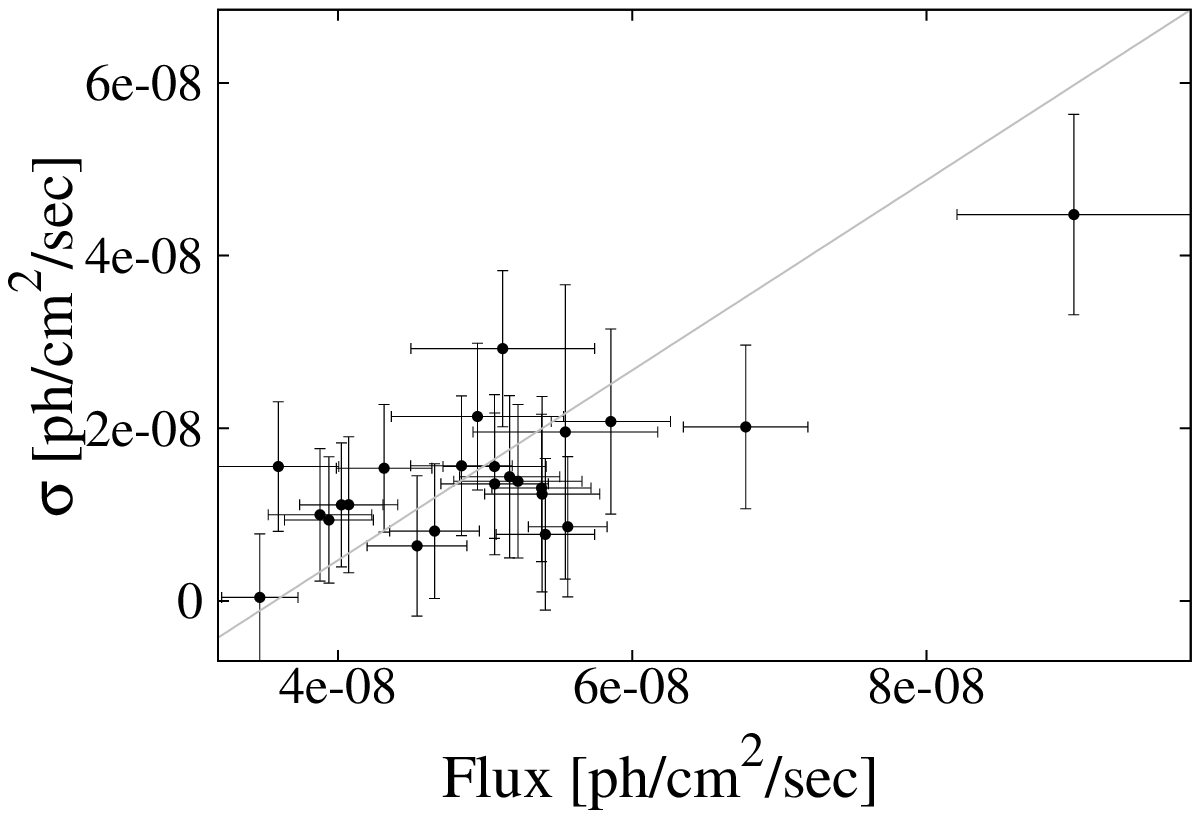}\label{Fig:excess}}
	\label{Fig:logn}
	\caption{Detection of lognormality in 1ES1011$+$496 in the Fermi energy band. The first panel shows the histogram of the observed fluxes (black points) fitted with a Gaussian (dotted blue line) and Lognormal (solid red line) function. A Lognormal fit is clearly preffered. The second panel shows the strong linear relationship seen between the flux and the excess rms. The black points denote data points averaged over 100days, and the solid gray line the linear fit.}
\end{figure*}

	Since for the present source the variablity is minimal in the MAXI and OVRO bands, and the sampling sparse in the UVOT and XRT bands, we restrict our study of lognormal flux variability to the Fermi band only. We fit the histogram of the observed fluxes with a Gaussian and a Lognormal function (Figure \ref{Fig:hist}), and find that a Lognormal fit (chi-sq/dof = 11/12) is statistically preffered over a Gaussian fit (chi-sq/dof $= \ 22.8/12$). We further plot the excess variance, $\sigma_{EXCESS}= \sqrt{S^2-\sigma^2_{err}}$ versus the mean flux in Figure \ref{Fig:excess}. The two parameters show a  strong linear correlation $r \ (prob)\ = \ 76\% \ (1.6e-5)$, and is well fit by a straight line of slope $1.17 \pm 0.18$ and a intercept of $(-3.9 \pm 0.8) \times 10^{-8}$ (chi-sq/dof $=\ 24.1/22$). This indicates a clear detection of lognormal temporal behaviour in this source.

\section[sect:SED]{Spectral Energy Distribution}\label{sect:SED}

The broadband SED of 1ES 1011+496, during different activity states, are modelled under the 
ambit of simple leptonic scenario. This, in turn, helps us to 
understand the the nature of the electron distribution responsible for the emission 
through synchrotron and SSC processes. We assume the electrons to be confined within a spherical 
zone of radius $R$ permeated by a tangled magnetic field $B$. As a result of the relativistic motion 
of the jet, the radiation is Doppler boosted along the line of sight. 
A good sampling of the SED from radio to $\gamma$-rays allows one to obtain a reasonable estimation 
of the physical parameters, under appropriate assumptions \citep{GMD,tavKN,sunder_3c279}. Notably, the smooth spectral curvatures, observed around the peak of the SED, may result from a convolution of the 
single particle emissivity with the assumed particle distribution. If the chosen particle distribution 
has a sharp break, then the observed curvature in the photon spectrum could be due to the emissivity 
function. On the other hand,
the underlying particle distribution itself can show a gradual transition causing the observed curvatures.
To investigate this, the observed SED is modelled with the following choices of particle distributions.

\begin{figure*}
	\centering
	\includegraphics[scale=0.8]{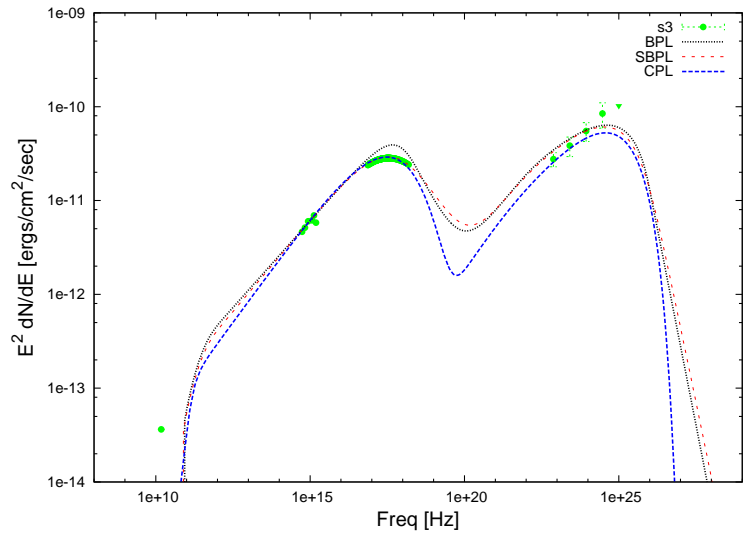}
	\caption{The state s3 modelled with the underlying spectrum as a BPL (dotted black line); a SBPL (dashed red line); and a CPL (dashed blue line). The BPL fails to reproduce the smooth curvature of the observed SED (shown in green).}
	\label{fig:s3_allps}
\end{figure*}

\begin{enumerate}[(i)]

	\item {\it Broken Power Law (BPL)}: In this case, we assume the electron spectrum to be  a  sharp broken power law with indices $p$ and $q$, given by

		\begin{equation}
				\label{eq:broken}
				N (\gamma) d\gamma =\left\{
				\begin{array}{ll}
					N_0 \gamma^{-p}d\gamma,&\mbox {~$\gamma_{min}<\gamma<\gamma_b$~} \\
					N_0 \gamma^{(q-p)}_b \gamma^{-q}d\gamma,&\mbox {~$\gamma_b<\gamma<\gamma_{max}$~}
				\end{array}
				\right.
		\end{equation}

	\item {\it Smooth Broken Power law (SBPL)}: Here, the electron distribution is a smooth broken power law with low energy index $p$ and the high 
		energy index $q$.
		\begin{equation}
			\label{eq:sbpo}
			N (\gamma) d\gamma = N_{0}  \frac{(\gamma_{b})^{-p}}{(\gamma/\gamma_{b})^p + (\gamma/\gamma_{b})^q} d\gamma ,\mbox {~$\gamma_{min}<\gamma<\gamma_{max}$~}
		\end{equation}

	\item {\it Power law with an exponential cutoff (CPL)}:  The particle distribution in this case is chosen to be a power law 
		with index $p$ and an exponentially decreasing tail, given by
	\begin{equation}
			\label{eq:cpo}
			N (\gamma) d\gamma = N_{0}  \gamma^{-p} exp\left(-\frac{\gamma}{\gamma_{c}}\right) d\gamma,\mbox {~$\gamma_{min}<\gamma<\gamma_{max}$~}
		\end{equation}
\end{enumerate}
Here, $\gamma_{min}$ and $\gamma_{max}$ are the minimum and maximum dimensionless energies
($E=\gamma mc^2$) of the
non-thermal electron distribution and $\gamma_b$ the
electron energy associated with the peak of the SED and $N_0$ the normalization.
To reduce the number of unknowns, the radius $R$ is fixed at $1.3 \times 10^{16}$cm, corresponding to a variability 
time scale of $t_{var} \approx 1 $ day (for the Doppler factor $\delta \approx 10$).  
In addition, the magnetic field energy density, $U_B \left(= \frac{B^2}{8 \pi} \right)$, is 
considered to be in near-equipartition with the particle 
energy density ($U_e$). The resultant model spectra, corresponding to epoch s3, for the above three 
choices of particle distribution are shown in Figure \ref{fig:s3_allps}, along with the observed fluxes. The 
governing physical parameters are given in Table \ref{partable}.  

We compare between the different fit models by incorporating the numerical SSC model into the XSPEC spectral fitting software to perform a $\chi^2$ minimization as followed in \cite{421_longterm}. {\it Swift}-XRT is binned to have 8 spectral points to avoid biasing the fit towards X-ray eneries. Since we are dealing with several different
instruments over a broad energy range, we assume model
systematics of 5\%. Our study shows that the commonly used electron spectrum, the BPL [e.g. \citet{GMD,kraw}], cannot 
explain the smooth curvature observed at the X-ray energies for this source implying that the synchrotron emissivity function alone is not sufficient to give rise to the observed curvature, and that the underlying particle spectrum itself must have a gradual transition as opposed to a sharp break. 
This suggests that the SBPL and CPL are the better choices to represent the observed SED and in Figure \ref{SED}
we show the model spectra corresponding to these particle distributions for all the three epochs considered in this study.
Both these models can well reproduce the observed spectrum, during all three epochs, and the absence 
of high energy X-ray/simultaneous TeV measurements prevents us from distinguishing between the two models.
Particularly, with the current sampling, the index $q$ and the $\gamma_{max}$ for the SBPL cannot 
be well constrained, and the latter is fixed at $10^7$. 
The model parameters describing the observed 
SED for the three epochs for the SBPL and the CPL are also listed in Table \ref{partable}. 

The different flux states can be reproduced by mainly changing the particle indices and the break energy; 
whereas, the variations in other parameters like the Doppler factor and the magnetic field are minimal. 
While the total bolometric luminosity, $L$, changes by more than a factor of $3$, the variations in $B$ and 
$\delta$ are less than $10\%$. This probably suggests, that the variation in the flux states may occur mainly 
due to changes in the underlying particle distributions, rather than the  other jet properties.

\section[sect:disc]{Discussions}\label{sect:disc}

The observations of lognormality in the long term (6 yrs) gamma ray flux distribution and the linear flux-rms relation imply that the $\gamma$-ray flux variability of  1ES~1011$+$496 is lognormal. Since similar trends have been seen in the X-ray band in sources like the Seyfert 1 galaxy, Mkn 766 \citep{Vaughan}, where the physical process responsible for the X-ray emission originates in the galactic disc, and other compact accreting systems like cataclysmic variables \citep{giannios}, such trends have been claimed as universal signs of accretion induced variability. The other option might be that the underlying parameters responsible for the observed emission (eg: the Doppler factor, magnetic field, etc) themselves have a lognormal time dependence \citep{giebels_lognorm}. Since the result of our spectral modelling indicates that the flux variability is mainly induced by changes in the particle spectrum rather than the other jet properties, it seems reasonable to believe that lognormal fluctuations in the accretion rate give rise to an injection rate into the jet with similar properties.

Moreover, a detailed study of the multiwavelength spectral behaviour of the source during three different epochs under simple synchrotron and SSC models demands an underlying electron distribution with a smooth curvature. Though such a requirement can be satisfied by assuming the underlying electron distribution as either SBPL or CPL, the absence of hard X-ray data prevents one from distinguishing between these two choices.  To interpret this, we consider a scenario where a non-thermal distribution of electrons $Q(\gamma)=Q_0\gamma^{-p}$  is continuously injected into a cooling region (CR) where they lose their energy through radiative processes as well as escape out at a rate defined by a characteristic time scale, $t_{esc}$.
The evolution of the particle number density, $N(\gamma,t)$, in the CR can then be conveniently described by the kinetic equation \citep{kardashev}
\begin{equation}
	\label{eq:KE}
	\frac{\partial N}{\partial t} - \frac{\partial}{\partial \gamma}(P(\gamma) N) + 
	\frac{N}{t_{esc}(\gamma)} = Q(\gamma) \Theta(\gamma-\gamma_{min}) \Theta(\gamma_{max}-\gamma)
\end{equation}
where $P(\gamma)$  is the energy loss rate due to synchrotron and SSC processes.
Assuming a power law dependence of escape timescale with energy, $t_{esc} = \tau \gamma ^{\xi}$, a semi analytical solution 
of  Equation \ref{eq:KE} can be attained when the loss processes are confined within the Thomson regime \citep{atoyan}.
However, detection of the source at VHE energies suggests that the Thomson scattering approximation of SSC process may not 
be valid and one needs to incorporate Klein-Nishina correction in the cross-section \citep{tavKN}. Hence, we numerically 
solve Equation \ref{eq:KE} using fully implicit finite difference scheme \citep{chang,Chiaberge}, while incorporating 
the exact Klein-Nishina cross-section for IC scattering \citep{blum}.

\begin{figure*}
	\centering
	\includegraphics[scale=0.8]{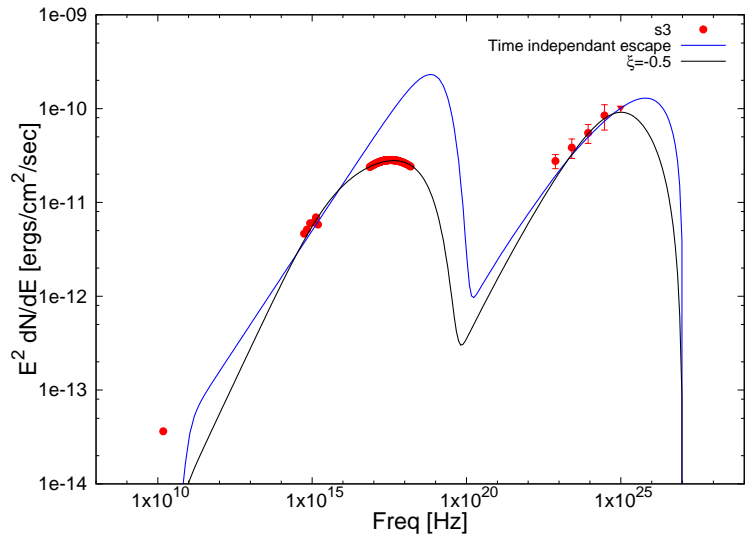}
	\caption{Model curves obtained by changing the energy dependence of the escape time scale for the state s3. The blue line is obtained for a energy independent escape ($\xi = 0$), and fails to reproduce the observed spectrum. The best match between data and model is obtained for $\xi=-0.5$ and is represented by the solid green line. The magnetic field $B$ is assumed to be $0.4G$ and the Doppler factor $\delta = 10$. The injected particle spectrum is a power of index $2.1$. }
	\label{s3_escape}
\end{figure*}

\begin{figure*}
	\centering
	\includegraphics[scale=0.8]{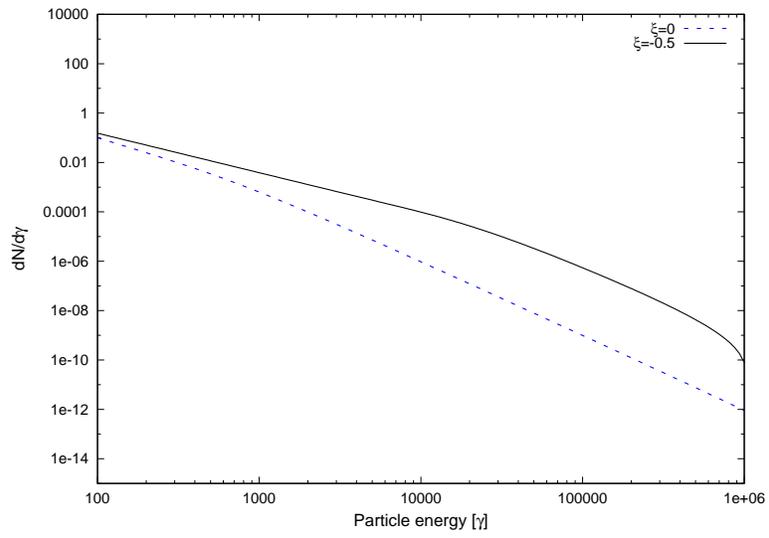}
	\caption{ Underlying particle spectra for the model curves shown in Figure \ref{s3_escape}. The dashed blue line is obtained for a energy independent escape ($\xi = 0$; which fails to reproduce the observed SED), and the solid black line corresponds to the particle spectrum which best models the observed SED ($\xi=-0.5$). } 
	\label{fig:partspec}
\end{figure*}

The case, $t_{esc}\rightarrow \infty$ (no escape), gives rise to a broken power law with the break occuring at the energy where the observation time is
equal to the cooling timescale of the particle; while  for the case $\xi = 0$ (constant $t_{esc}$), a steady state broken power-law particle distribution is
eventually attained where the break corresponds to the particle energy at which the escape timescale equats to its cooling timescale.
However, in these cases, the spectral tranistion at the break energy is too sharp to reproduce the observed SED and a gradual transition can be 
acheived by considering $\xi \neq 0$. We found that the smooth spectral curvature demanded by the observation can be attained by fixing 
$\xi\approx 0.5$. In Figure \ref{s3_escape}, we show the resultant model SED corresponding to $\xi = 0$ (blue line) and $\xi=0.5$ along with 
the observed fluxes for the state s3. The underlying particle distribution corresponding to these values of $\xi$ is shown in Figure \ref{fig:partspec}.

 Alternate to this interpretation, a smooth curvature in the particle distribution can also be a result of time averaging of an
evolving particle distribution. For instance, an episodic injection of a power-law particle distribution into CR can cause the high energy 
cut-off to shift towards lower energy with time which will be reflected as a smooth curvature at high energy in the time averaged spectrum. Also, an episodic injection with an energy-dependent escape gives rise to a particle distribution similar to a cutoff powr-law. However,
this interpretation fails to explain the observed variability of the gamma ray flare since the observed cooling timescale of the GeV gamma-ray emitting 
electrons will come out to be \citep{pankaj1222} 

\begin{align}
	t_{\rm cool,GeV} &\approx 3\times10^{10} B^{-7/4} \left(\frac{\delta}{\nu_{\rm GeV}}\right)^{1/4} \, {\rm sec}\\
   & \approx 0.5 \, {\rm days}
\end{align}
Here, the cooling timescale is estimated for the parameters provided in Table \ref{partable} and $\nu_{\rm GeV}$ corresponds to the frequency of the 
gamma ray photon falling on the Compton peak. This time is much smaller than the observed flare duration of $\approx 30 {\rm days}$ as seen in Figure \ref{fig:lc}. Similarly,
the particle spectrum injected into the CR itself can show smooth curvatures due to an underlying complex particle acceleration proces. 
For example, an energy dependent acceleration process is known to give rise to significant curvature in the accelerated particle
distribution \citep{massaro,zirak}. However, in this present work, we only consider a simplistic scenario where the smooth curvature can
be introduced by considering an energy dependent escape from the main emission region.

\section[sect:conclusion]{Conclusions}\label{sect:conclusion}

The blazar, 1ES\,1011+496, underwent a major $\gamma$-ray flare during February, 2014, triggering observations 
at other wavebands, thereby, providing simultaneous observations of the source at radio/optical/X-ray/$\gamma$-ray energies.
The TeV flare seen by VERITAS between February 3-11, 2014 decayed down by February 19, 2014 as indicated 
by the HAGAR upper limits. This is also seen in the flux decrease in the Fermi and the X-ray bands. 
In this work, we analyzed the simultaneous multiwavelength spectrum of the source and obtained the broadband SED during 
three different epochs. The observed SEDs over these epochs clearly show a trend of the synchrotron peak moving to 
higher energies during increased flux states, similar to the "bluer when brighter" trend seen 
by \cite{bottcher_1011} during 5 years of optical observations of this source. However, \cite{bottcher_1011} attributed the 
observed variability primarily due to magnetic field changes in the jet, whereas our broadband spectral modelling results 
indicate the spectral behaviour is dominated by changes in the particle spectrum. 

The spectra during all the states demand a smooth curvature in the underlying particle spectrum, eg, a SBPL or a CPL. We show that such a smoothly varying particle spectrum, can be easily obtained by assuming an energy dependent particle escape ($t_{esc} \propto \gamma^{\xi}$) mechanism in the jet.  The inferred injected particle power law index $p \sim 2.1 $ indicates that the particles are most likely accelerated at relativistic shocks \citep{sironi}. However, the non availability of hard X-ray observation presently prevents us from
distinguishing between CPL and SBPL particle spectra. Future observations in the hard X-ray band from the newly launched 
ASTROSAT \citep{astrosat}, can be crucial in resolving this uncertainty.

The detection of lognormal flux variability in this source follows similar recent detections in other blazars. While the $\gamma$-ray flux distribution could be modelled by a single lognormal distribution for other HBLs (Mkn421; \cite{421_longterm} and PKS2155-304; \cite{lognorm_2155}), the FSRQ PKS~1510-089 \citep{pankaj_ln} required a sum of two such distributions. With the Fermi mission now into its ninth year of operation, we have unprecedented continuous flux measurements for a large sample of blazars. A systematic study of the same can throw new light on the origin of the jet launching mechanisms in supermassive blackholes.

\section{Acknowledgements}
We thank the scientific and technical personnel of HAGAR groups at IIA, Bengaluru and TIFR, Mumbai for observations with HAGAR system.
This research has made use of data,
software and/or web tools obtained from NASAs High Energy Astrophysics Science Archive
Research Center (HEASARC), a service of Goddard Space Flight Center and the Smithsonian
Astrophysical Observatory. Part of this work is based on archival data, software, or online
services provided by the ASI Science Data Center (ASDC).	This research has made use of the
  XRT Data Analysis Software (XRTDAS) developed under the responsibility
the ASI Science Data Center (ASDC), Italy, and the  NuSTAR Data Analysis Software (NuSTARDAS) jointly developed by the ASI Science
    Data Center (ASDC, Italy) and the California Institute of Technology (Caltech, USA). 
The OVRO 40 M Telescope Fermi Blazar Monitoring Program is supported by NASA under awards 
NNX08AW31G and NNX11A043G, and by the NSF under awards AST-0808050 and AST-1109911. We are grateful to the anonymous referee for the constructive criticism and helpful comments which lead to a major change in the discussions section.

\bibliography{1es1011,hagar,Mkn421}

\begin{thebibliography}{}
\expandafter\ifx\csname natexlab\endcsname\relax\def\natexlab#1{#1}\fi

\bibitem[{{Abdo} {et~al.}(2010){Abdo}, {Ackermann}, {Agudo}, {Ajello}, {Aller},
  {Aller}, {Angelakis}, {Arkharov}, {Axelsson}, {Bach}, \& et~al.}]{Abdo_SED}
{Abdo}, A.~A., {Ackermann}, M., {Agudo}, I., {et~al.} 2010, \apj, 716, 30

\bibitem[{{Acero} {et~al.}(2015){Acero}, {Ackermann}, {Ajello}, {Albert},
  {Atwood}, {Axelsson}, {Baldini}, {Ballet}, {Barbiellini}, {Bastieri},
  {Belfiore}, {Bellazzini}, {Bissaldi}, {Blandford}, {Bloom}, {Bogart},
  {Bonino}, {Bottacini}, {Bregeon}, {Britto}, {Bruel}, {Buehler}, {Burnett},
  {Buson}, {Caliandro}, {Cameron}, {Caputo}, {Caragiulo}, {Caraveo},
  {Casandjian}, {Cavazzuti}, {Charles}, {Chaves}, {Chekhtman}, {Cheung},
  {Chiang}, {Chiaro}, {Ciprini}, {Claus}, {Cohen-Tanugi}, {Cominsky}, {Conrad},
  {Cutini}, {DAmmando}, {de Angelis}, {DeKlotz}, {de Palma}, {Desiante},
  {Digel}, {Di Venere}, {Drell}, {Dubois}, {Dumora}, {Favuzzi}, {Fegan},
  {Ferrara}, {Finke}, {Franckowiak}, {Fukazawa}, {Funk}, {Fusco}, {Gargano},
  {Gasparrini}, {Giebels}, {Giglietto}, {Giommi}, {Giordano}, {Giroletti},
  {Glanzman}, {Godfrey}, {Grenier}, {Grondin}, {Grove}, {Guillemot}, {Guiriec},
  {Hadasch}, {Harding}, {Hays}, {Hewitt}, {Hill}, {Horan}, {Iafrate}, {Jogler},
  {J{\'o}hannesson}, {Johnson}, {Johnson}, {Johnson}, {Johnson}, {Kamae},
  {Kataoka}, {Katsuta}, {Kuss}, {La Mura}, {Landriu}, {Larsson}, {Latronico},
  {Lemoine-Goumard}, {Li}, {Li}, {Longo}, {Loparco}, {Lott}, {Lovellette},
  {Lubrano}, {Madejski}, {Massaro}, {Mayer}, {Mazziotta}, {McEnery},
  {Michelson}, {Mirabal}, {Mizuno}, {Moiseev}, {Mongelli}, {Monzani},
  {Morselli}, {Moskalenko}, {Murgia}, {Nuss}, {Ohno}, {Ohsugi}, {Omodei},
  {Orienti}, {Orlando}, {Ormes}, {Paneque}, {Panetta}, {Perkins},
  {Pesce-Rollins}, {Piron}, {Pivato}, {Porter}, {Racusin}, {Rando}, {Razzano},
  {Razzaque}, {Reimer}, {Reimer}, {Reposeur}, {Rochester}, {Romani},
  {Salvetti}, {S{\'a}nchez-Conde}, {Saz Parkinson}, {Schulz}, {Siskind},
  {Smith}, {Spada}, {Spandre}, {Spinelli}, {Stephens}, {Strong}, {Suson},
  {Takahashi}, {Takahashi}, {Tanaka}, {Thayer}, {Thayer}, {Thompson},
  {Tibaldo}, {Tibolla}, {Torres}, {Torresi}, {Tosti}, {Troja}, {Van Klaveren},
  {Vianello}, {Winer}, {Wood}, {Wood}, \& {Zimmer}}]{3FGL}
{Acero}, F., {Ackermann}, M., {Ajello}, M., {et~al.} 2015, \apjs, 218, 23

\bibitem[{{Ahnen} {et~al.}(2016{\natexlab{a}}){Ahnen}, {Ansoldi}, {Antonelli},
  {Antoranz}, {Babic}, {Banerjee}, {Bangale}, {Barres de Almeida}, {Barrio},
  {Becerra Gonz{\'a}lez}, {Bednarek}, {Bernardini}, {Biasuzzi}, {Biland},
  {Blanch}, {Bonnefoy}, {Bonnoli}, {Borracci}, {Bretz}, {Carmona}, {Carosi},
  {Chatterjee}, {Clavero}, {Colin}, {Colombo}, {Contreras}, {Cortina},
  {Covino}, {Da Vela}, {Dazzi}, {De Angelis}, {De Lotto}, {de O{\~n}a
  Wilhelmi}, {Delgado Mendez}, {Di Pierro}, {Dominis Prester}, {Dorner},
  {Doro}, {Einecke}, {Eisenacher Glawion}, {Elsaesser}, {Fern{\'a}ndez-Barral},
  {Fidalgo}, {Fonseca}, {Font}, {Frantzen}, {Fruck}, {Galindo}, {Garc{\'{\i}}a
  L{\'o}pez}, {Garczarczyk}, {Garrido Terrats}, {Gaug}, {Giammaria},
  {Godinovi{\'c}}, {Gonz{\'a}lez Mu{\~n}oz}, {Guberman}, {Hahn}, {Hanabata},
  {Hayashida}, {Herrera}, {Hose}, {Hrupec}, {Hughes}, {Idec}, {Kodani},
  {Konno}, {Kubo}, {Kushida}, {La Barbera}, {Lelas}, {Lindfors}, {Lombardi},
  {Longo}, {L{\'o}pez}, {L{\'o}pez-Coto}, {L{\'o}pez-Oramas}, {Lorenz},
  {Majumdar}, {Makariev}, {Mallot}, {Maneva}, {Manganaro}, {Mannheim},
  {Maraschi}, {Marcote}, {Mariotti}, {Mart{\'{\i}}nez}, {Mazin}, {Menzel},
  {Miranda}, {Mirzoyan}, {Moralejo}, {Moretti}, {Nakajima}, {Neustroev},
  {Niedzwiecki}, {Nievas Rosillo}, {Nilsson}, {Nishijima}, {Noda}, {Orito},
  {Overkemping}, {Paiano}, {Palacio}, {Palatiello}, {Paneque}, {Paoletti},
  {Paredes}, {Paredes-Fortuny}, {Persic}, {Poutanen}, {Prada Moroni},
  {Prandini}, {Puljak}, {Rhode}, {Rib{\'o}}, {Rico}, {Rodriguez Garcia},
  {Saito}, {Satalecka}, {Schultz}, {Schweizer}, {Shore}, {Sillanp{\"a}{\"a}},
  {Sitarek}, {Snidaric}, {Sobczynska}, {Stamerra}, {Steinbring}, {Strzys},
  {Takalo}, {Takami}, {Tavecchio}, {Temnikov}, {Terzi{\'c}}, {Tescaro},
  {Teshima}, {Thaele}, {Torres}, {Toyama}, {Treves}, {Verguilov}, {Vovk},
  {Ward}, {Will}, {Wu}, \& {Zanin}}]{1011_2014}
{Ahnen}, M.~L., {Ansoldi}, S., {Antonelli}, L.~A., {et~al.} 2016{\natexlab{a}},
  \aap, 590, A24

\bibitem[{{Ahnen} {et~al.}(2016{\natexlab{b}}){Ahnen}, {Ansoldi}, {Antonelli},
  {Antoranz}, {Babic}, {Banerjee}, {Bangale}, {de Almeida}, {Barrio},
  {Gonz{\'a}lez}, {Bednarek}, {Bernardini}, {Biasuzzi}, {Biland}, {Blanch},
  {Bonnefoy}, {Bonnoli}, {Borracci}, {Bretz}, {Carmona}, {Carosi},
  {Chatterjee}, {Clavero}, {Colin}, {Colombo}, {Contreras}, {Cortina},
  {Covino}, {Da Vela}, {Dazzi}, {De Angelis}, {De Caneva}, {De Lotto}, {de
  O{\~n}a Wilhelmi}, {Mendez}, {Pierro}, {Prester}, {Dorner}, {Doro},
  {Einecke}, {Elsaesser}, {Fern{\'a}ndez-Barral}, {Fidalgo}, {Fonseca}, {Font},
  {Frantzen}, {Fruck}, {Galindo}, {L{\'o}pez}, {Garczarczyk}, {Terrats},
  {Gaug}, {Giammaria}, {(Eisenacher)}, {Godinovi{\'c}}, {Mu{\~n}oz},
  {Guberman}, {Hanabata}, {Hayashida}, {Herrera}, {Hose}, {Hrupec}, {Hughes},
  {Idec}, {Kodani}, {Konno}, {Kubo}, {Kushida}, {Barbera}, {Lelas}, {Lindfors},
  {Lombardi}, {Longo}, {L{\'o}pez}, {L{\'o}pez-Coto}, {L{\'o}pez-Oramas},
  {Lorenz}, {Majumdar}, {Makariev}, {Mallot}, {Maneva}, {Manganaro},
  {Mannheim}, {Maraschi}, {Marcote}, {Mariotti}, {Mart{\'{\i}}nez}, {Mazin},
  {Menzel}, {Miranda}, {Mirzoyan}, {Moralejo}, {Nakajima}, {Neustroev},
  {Niedzwiecki}, {Rosillo}, {Nilsson}, {Nishijima}, {Noda}, {Orito},
  {Overkemping}, {Paiano}, {Palacio}, {Palatiello}, {Paneque}, {Paoletti},
  {Paredes}, {Paredes-Fortuny}, {Persic}, {Poutanen}, {Moroni}, {Prandini},
  {Puljak}, {Reinthal}, {Rhode}, {Rib{\'o}}, {Rico}, {Garcia}, {R{\"u}gamer},
  {Saito}, {Satalecka}, {Scapin}, {Schultz}, {Schweizer}, {Shore},
  {Sillanp{\"a}{\"a}}, {Sitarek}, {Snidaric}, {Sobczynska}, {Stamerra},
  {Steinbring}, {Strzys}, {Takalo}, {Takami}, {Tavecchio}, {Temnikov},
  {Terzi{\'c}}, {Tescaro}, {Teshima}, {Thaele}, {Torres}, {Toyama}, {Treves},
  {Verguilov}, {Vovk}, {Ward}, {Will}, {Wu}, {Zanin}, {Lucarelli}, {Pittori},
  {Vercellone}, {Berdyugin}, {Carini}, {L{\"a}hteenm{\"a}ki}, {Pasanen},
  {Pease}, {Sainio}, {Tornikoski}, \& {Walters}}]{1011_2008}
---. 2016{\natexlab{b}}, \mnras, 459, 2286

\bibitem[{{Albert} {et~al.}(2007){Albert}, {Aliu}, {Anderhub}, {Antoranz},
  {Armada}, {Baixeras}, {Barrio}, {Bartko}, {Bastieri}, {Becker}, {Bednarek},
  {Berger}, {Bigongiari}, {Biland}, {Bock}, {Bordas}, {Bosch-Ramon}, {Bretz},
  {Britvitch}, {Camara}, {Carmona}, {Chilingarian}, {Coarasa}, {Commichau},
  {Contreras}, {Cortina}, {Costado}, {Curtef}, {Danielyan}, {Dazzi}, {De
  Angelis}, {Delgado}, {de los Reyes}, {De Lotto}, {Domingo-Santamar{\'{\i}}a},
  {Dorner}, {Doro}, {Errando}, {Fagiolini}, {Ferenc}, {Fern{\'a}ndez}, {Firpo},
  {Flix}, {Fonseca}, {Font}, {Fuchs}, {Galante}, {Garc{\'{\i}}a-L{\'o}pez},
  {Garczarczyk}, {Gaug}, {Giller}, {Goebel}, {Hakobyan}, {Hayashida},
  {Hengstebeck}, {Herrero}, {H{\"o}hne}, {Hose}, {Hsu}, {Jacon}, {Jogler},
  {Kosyra}, {Kranich}, {Kritzer}, {Laille}, {Lindfors}, {Lombardi}, {Longo},
  {L{\'o}pez}, {L{\'o}pez}, {Lorenz}, {Majumdar}, {Maneva}, {Mannheim},
  {Mansutti}, {Mariotti}, {Mart{\'{\i}}nez}, {Mazin}, {Merck}, {Meucci},
  {Meyer}, {Miranda}, {Mirzoyan}, {Mizobuchi}, {Moralejo}, {Nieto}, {Nilsson},
  {Ninkovic}, {O{\~n}a-Wilhelmi}, {Otte}, {Oya}, {Paneque}, {Panniello},
  {Paoletti}, {Paredes}, {Pasanen}, {Pascoli}, {Pauss}, {Pegna}, {Perlman},
  {Persic}, {Peruzzo}, {Piccioli}, {Prandini}, {Puchades}, {Raymers}, {Rhode},
  {Rib{\'o}}, {Rico}, {Rissi}, {Robert}, {R{\"u}gamer}, {Saggion}, {Saito},
  {S{\'a}nchez}, {Sartori}, {Scalzotto}, {Scapin}, {Schmitt}, {Schweizer},
  {Shayduk}, {Shinozaki}, {Shore}, {Sidro}, {Sillanp{\"a}{\"a}}, {Sobczynska},
  {Stamerra}, {Stark}, {Takalo}, {Tavecchio}, {Temnikov}, {Tescaro}, {Teshima},
  {Torres}, {Turini}, {Vankov}, {Vitale}, {Wagner}, {Wibig}, {Wittek},
  {Zandanel}, {Zanin}, \& {Zapatero}}]{1es1011_Magic}
{Albert}, J., {Aliu}, E., {Anderhub}, H., {et~al.} 2007, \apjl, 667, L21

\bibitem[{{Aleksi{\'c}} {et~al.}(2016){Aleksi{\'c}}, {Ansoldi}, {Antonelli},
  {Antoranz}, {Arcaro}, {Babic}, {Bangale}, {Barres de Almeida}, {Barrio},
  {Becerra Gonz{\'a}lez}, {Bednarek}, {Bernardini}, {Biasuzzi}, {Biland},
  {Blanch}, {Bonnefoy}, {Bonnoli}, {Borracci}, {Bretz}, {Carmona}, {Carosi},
  {Colin}, {Colombo}, {Contreras}, {Cortina}, {Covino}, {Da Vela}, {Dazzi}, {De
  Angelis}, {De Caneva}, {De Lotto}, {de O{\~n}a Wilhelmi}, {Delgado Mendez},
  {Di Pierro}, {Dominis Prester}, {Dorner}, {Doro}, {Einecke}, {Eisenacher},
  {Elsaesser}, {Fern{\'a}ndez-Barral}, {Fidalgo}, {Fonseca}, {Font},
  {Frantzen}, {Fruck}, {Galindo}, {Garc{\'{\i}}a L{\'o}pez}, {Garczarczyk},
  {Garrido Terrats}, {Gaug}, {Godinovi{\'c}}, {Gonz{\'a}lez Mu{\~n}oz},
  {Gozzini}, {Hadasch}, {Hanabata}, {Hayashida}, {Herrera}, {Hose}, {Hrupec},
  {Idec}, {Kadenius}, {Kellermann}, {Knoetig}, {Kodani}, {Konno}, {Krause},
  {Kubo}, {Kushida}, {La Barbera}, {Lelas}, {Lewandowska}, {Lindfors},
  {Lombardi}, {Longo}, {L{\'o}pez}, {L{\'o}pez-Coto}, {L{\'o}pez-Oramas},
  {Lorenz}, {Lozano}, {Makariev}, {Mallot}, {Maneva}, {Mannheim}, {Maraschi},
  {Marcote}, {Mariotti}, {Mart{\'{\i}}nez}, {Mazin}, {Menzel}, {Miranda},
  {Mirzoyan}, {Moralejo}, {Munar-Adrover}, {Nakajima}, {Neustroev},
  {Niedzwiecki}, {Nievas Rosillo}, {Nilsson}, {Nishijima}, {Noda}, {Orito},
  {Overkemping}, {Paiano}, {Palatiello}, {Paneque}, {Paoletti}, {Paredes},
  {Paredes-Fortuny}, {Persic}, {Poutanen}, {Prada Moroni}, {Prandini},
  {Puljak}, {Reinthal}, {Rhode}, {Rib{\'o}}, {Rico}, {Rodriguez Garcia},
  {Saito}, {Saito}, {Satalecka}, {Scalzotto}, {Scapin}, {Schweizer}, {Shore},
  {Sillanp{\"a}{\"a}}, {Sitarek}, {Snidaric}, {Sobczynska}, {Stamerra},
  {Steinbring}, {Strzys}, {Takalo}, {Takami}, {Tavecchio}, {Temnikov},
  {Terzi{\'c}}, {Tescaro}, {Teshima}, {Thaele}, {Torres}, {Toyama}, {Treves},
  {Vogler}, {Will}, {Zanin}, {Buson}, {D'Ammando}, {L{\"a}hteenm{\"a}ki},
  {Hovatta}, {Kovalev}, {Lister}, {Max-Moerbeck}, {Mundell}, {Pushkarev},
  {Rastorgueva-Foi}, {Readhead}, {Richards}, {Tammi}, {Sanchez}, {Tornikoski},
  {Savolainen}, \& {Steele}}]{1011_2011-12}
{Aleksi{\'c}}, J., {Ansoldi}, S., {Antonelli}, L.~A., {et~al.} 2016, \aap, 591,
  A10

\bibitem[{{Atoyan} \& {Aharonian}(1999)}]{atoyan}
{Atoyan}, A.~M., \& {Aharonian}, F.~A. 1999, \mnras, 302, 253

\bibitem[{{B{\l}a{\.z}ejowski} {et~al.}(2000){B{\l}a{\.z}ejowski}, {Sikora},
  {Moderski}, \& {Madejski}}]{leptonic_bla}
{B{\l}a{\.z}ejowski}, M., {Sikora}, M., {Moderski}, R., \& {Madejski}, G.~M.
  2000, \apj, 545, 107

\bibitem[{{Bloom} \& {Marscher}(1996)}]{leptonic_bloom}
{Bloom}, S.~D., \& {Marscher}, A.~P. 1996, \apj, 461, 657

\bibitem[{{Blumenthal} \& {Gould}(1970)}]{blum}
{Blumenthal}, G.~R., \& {Gould}, R.~J. 1970, Reviews of Modern Physics, 42, 237

\bibitem[{{B{\"o}ttcher}(2007)}]{bottcher2007}
{B{\"o}ttcher}, M. 2007, \apss, 309, 95

\bibitem[{{B{\"o}ttcher} {et~al.}(2010){B{\"o}ttcher}, {Hivick}, {Dashti},
  {Fultz}, {Gupta}, {Gusbar}, {Joshi}, {Lamerato}, {Peery}, {Principe},
  {Rajasingam}, {Roustazadeh}, \& {Shields}}]{bottcher_1011}
{B{\"o}ttcher}, M., {Hivick}, B., {Dashti}, J., {et~al.} 2010, \apj, 725, 2344

\bibitem[{{Burrows} {et~al.}(2005){Burrows}, {Hill}, {Nousek}, {Kennea},
  {Wells}, {Osborne}, {Abbey}, {Beardmore}, {Mukerjee}, {Short}, {Chincarini},
  {Campana}, {Citterio}, {Moretti}, {Pagani}, {Tagliaferri}, {Giommi},
  {Capalbi}, {Tamburelli}, {Angelini}, {Cusumano}, {Br{\"a}uninger}, {Burkert},
  \& {Hartner}}]{XRT_instru}
{Burrows}, D.~N., {Hill}, J.~E., {Nousek}, J.~A., {et~al.} 2005, \ssr, 120, 165

\bibitem[{{Cerruti}(2015)}]{VERITAS_update}
{Cerruti}, M. 2015, ArXiv e-prints, arXiv:1501.03554

\bibitem[{{Chang} \& {Cooper}(1970)}]{chang}
{Chang}, J.~S., \& {Cooper}, G. 1970, Journal of Computational Physics, 6, 1

\bibitem[{{Chevalier} {et~al.}(2015){Chevalier}, {Kastendieck}, {Rieger},
  {Maurin}, {Lenain}, \& {Giovanni Lamanna for the
  H.~E.~S.~S.~collaboration}}]{lognorm_2155}
{Chevalier}, J., {Kastendieck}, M.~A., {Rieger}, F., {et~al.} 2015, ArXiv
  e-prints, arXiv:1509.03104

\bibitem[{{Chiaberge} \& {Ghisellini}(1999)}]{Chiaberge}
{Chiaberge}, M., \& {Ghisellini}, G. 1999, \mnras, 306, 551

\bibitem[{{Chitnis} {et~al.}(2009){Chitnis}, {Pendharkar}, {Bose}, {Agrawal},
  {Rao}, \& {Misra}}]{varsha_var}
{Chitnis}, V.~R., {Pendharkar}, J.~K., {Bose}, D., {et~al.} 2009, \apj, 698,
  1207

\bibitem[{{Corbet} \& {Shrader}(2014)}]{ATEL5888}
{Corbet}, R.~H.~D., \& {Shrader}, C.~R. 2014, The Astronomer's Telegram, 5888,
  1

\bibitem[{{Costamante} \& {Ghisellini}(2002)}]{costamante}
{Costamante}, L., \& {Ghisellini}, G. 2002, \aap, 384, 56

\bibitem[{{Dermer} \& {Schlickeiser}(1993)}]{leptonic_dermer}
{Dermer}, C.~D., \& {Schlickeiser}, R. 1993, \apj, 416, 458

\bibitem[{{Ghisellini} {et~al.}(1996){Ghisellini}, {Maraschi}, \&
  {Dondi}}]{GMD}
{Ghisellini}, G., {Maraschi}, L., \& {Dondi}, L. 1996, \aaps, 120, C503

\bibitem[{{Giannios}(2013)}]{giannios}
{Giannios}, D. 2013, \mnras, 431, 355

\bibitem[{{Giebels} \& {Degrange}(2009)}]{giebels_lognorm}
{Giebels}, B., \& {Degrange}, B. 2009, \aap, 503, 797

\bibitem[{{Gothe} {et~al.}(2013){Gothe}, {Prabhu}, {Vishwanath}, {Acharya},
  {Srinivasan}, {Chitnis}, {Kamath}, {Srinivasulu}, {Saleem}, {Kemkar},
  {Mahesh}, {Gabriel}, {Manoharan}, {Dorji}, {Dorjai}, {Angchuk}, {D'souza},
  {Duhan}, {Nagesh}, {Rao}, {Sharma}, {Singh}, {Sudersanan}, {Thsering},
  {Upadhya}, {Anupama}, {Britto}, {Cowsik}, {Saha}, \& {Shukla}}]{kiranHAGAR}
{Gothe}, K.~S., {Prabhu}, T.~P., {Vishwanath}, P.~R., {et~al.} 2013,
  Experimental Astronomy, 35, 489

\bibitem[{{Hartman} {et~al.}(1999){Hartman}, {Bertsch}, {Bloom}, {Chen},
  {Deines-Jones}, {Esposito}, {Fichtel}, {Friedlander}, {Hunter}, {McDonald},
  {Sreekumar}, {Thompson}, {Jones}, {Lin}, {Michelson}, {Nolan}, {Tompkins},
  {Kanbach}, {Mayer-Hasselwander}, {M{\"u}cke}, {Pohl}, {Reimer}, {Kniffen},
  {Schneid}, {von Montigny}, {Mukherjee}, \& {Dingus}}]{egret3C}
{Hartman}, R.~C., {Bertsch}, D.~L., {Bloom}, S.~D., {et~al.} 1999, \apjs, 123,
  79

\bibitem[{{Kalberla} {et~al.}(2005){Kalberla}, {Burton}, {Hartmann}, {Arnal},
  {Bajaja}, {Morras}, \& {P{\"o}ppel}}]{LAB}
{Kalberla}, P.~M.~W., {Burton}, W.~B., {Hartmann}, D., {et~al.} 2005, \aap,
  440, 775

\bibitem[{{Kapanadze}(2014)}]{ATEL5866}
{Kapanadze}, B. 2014, The Astronomer's Telegram, 5866, 1

\bibitem[{{Kardashev}(1962)}]{kardashev}
{Kardashev}, N.~S. 1962, \sovast, 6, 317

\bibitem[{{Kneiske} \& {Dole}(2010)}]{kneiske}
{Kneiske}, T.~M., \& {Dole}, H. 2010, \aap, 515, A19

\bibitem[{{Krawczynski} {et~al.}(2004){Krawczynski}, {Hughes}, {Horan},
  {Aharonian}, {Aller}, {Aller}, {Boltwood}, {Buckley}, {Coppi}, {Fossati},
  {G{\"o}tting}, {Holder}, {Horns}, {Kurtanidze}, {Marscher}, {Nikolashvili},
  {Remillard}, {Sadun}, \& {Schr{\"o}der}}]{kraw}
{Krawczynski}, H., {Hughes}, S.~B., {Horan}, D., {et~al.} 2004, \apj, 601, 151

\bibitem[{{Kushwaha} {et~al.}(2016){Kushwaha}, {Chandra}, {Misra},
  {Sahayanathan}, {Singh}, \& {Baliyan}}]{pankaj_ln}
{Kushwaha}, P., {Chandra}, S., {Misra}, R., {et~al.} 2016, \apjl, 822, L13

\bibitem[{{Kushwaha} {et~al.}(2014){Kushwaha}, {Sahayanathan}, {Lekshmi},
  {Singh}, {Bhattacharyya}, \& {Bhattacharya}}]{pankaj1222}
{Kushwaha}, P., {Sahayanathan}, S., {Lekshmi}, R., {et~al.} 2014, \mnras, 442,
  131

\bibitem[{{Mannheim} \& {Biermann}(1992)}]{had1}
{Mannheim}, K., \& {Biermann}, P.~L. 1992, \aap, 253, L21

\bibitem[{{Maraschi} {et~al.}(1992){Maraschi}, {Ghisellini}, \&
  {Celotti}}]{leptonic_maraschi}
{Maraschi}, L., {Ghisellini}, G., \& {Celotti}, A. 1992, \apjl, 397, L5

\bibitem[{{Massaro} {et~al.}(2004){Massaro}, {Perri}, {Giommi}, \&
  {Nesci}}]{massaro}
{Massaro}, E., {Perri}, M., {Giommi}, P., \& {Nesci}, R. 2004, \aap, 413, 489

\bibitem[{{Matsuoka} {et~al.}(2009){Matsuoka}, {Kawasaki}, {Ueno}, {Tomida},
  {Kohama}, {Suzuki}, {Adachi}, {Ishikawa}, {Mihara}, {Sugizaki}, {Isobe},
  {Nakagawa}, {Tsunemi}, {Miyata}, {Kawai}, {Kataoka}, {Morii}, {Yoshida},
  {Negoro}, {Nakajima}, {Ueda}, {Chujo}, {Yamaoka}, {Yamazaki}, {Nakahira},
  {You}, {Ishiwata}, {Miyoshi}, {Eguchi}, {Hiroi}, {Katayama}, \&
  {Ebisawa}}]{MAXI}
{Matsuoka}, M., {Kawasaki}, K., {Ueno}, S., {et~al.} 2009, \pasj, 61, 999

\bibitem[{{McHardy}(2008)}]{blazar_var}
{McHardy}, I. 2008, in Blazar Variability across the Electromagnetic Spectrum,
  14

\bibitem[{{Mirzoyan}(2014)}]{ATEL5887}
{Mirzoyan}, R. 2014, The Astronomer's Telegram, 5887, 1

\bibitem[{{Moretti} {et~al.}(2005){Moretti}, {Campana}, {Mineo}, {Romano},
  {Abbey}, {Angelini}, {Beardmore}, {Burkert}, {Burrows}, {Capalbi},
  {Chincarini}, {Citterio}, {Cusumano}, {Freyberg}, {Giommi}, {Goad}, {Godet},
  {Hartner}, {Hill}, {Kennea}, {La Parola}, {Mangano}, {Morris}, {Nousek},
  {Osborne}, {Page}, {Pagani}, {Perri}, {Tagliaferri}, {Tamburelli}, \&
  {Wells}}]{XRTPSF}
{Moretti}, A., {Campana}, S., {Mineo}, T., {et~al.} 2005, in Society of
  Photo-Optical Instrumentation Engineers (SPIE) Conference Series, Vol. 5898,
  UV, X-Ray, and Gamma-Ray Space Instrumentation for Astronomy XIV, ed.
  O.~H.~W. {Siegmund}, 360--368

\bibitem[{{M{\"u}cke} \& {Protheroe}(2001)}]{had2}
{M{\"u}cke}, A., \& {Protheroe}, R.~J. 2001, Astroparticle Physics, 15, 121

\bibitem[{{M{\"u}cke} {et~al.}(2003){M{\"u}cke}, {Protheroe}, {Engel},
  {Rachen}, \& {Stanev}}]{had3}
{M{\"u}cke}, A., {Protheroe}, R.~J., {Engel}, R., {Rachen}, J.~P., \& {Stanev},
  T. 2003, Astroparticle Physics, 18, 593

\bibitem[{{Nilsson} {et~al.}(2007){Nilsson}, {Pasanen}, {Takalo}, {Lindfors},
  {Berdyugin}, {Ciprini}, \& {Pforr}}]{host_corr}
{Nilsson}, K., {Pasanen}, M., {Takalo}, L.~O., {et~al.} 2007, \aap, 475, 199

\bibitem[{{Padovani} \& {Giommi}(1995)}]{PadGio}
{Padovani}, P., \& {Giommi}, P. 1995, \apj, 444, 567

\bibitem[{{Richards} {et~al.}(2011){Richards}, {Max-Moerbeck}, {Pavlidou},
  {King}, {Pearson}, {Readhead}, {Reeves}, {Shepherd}, {Stevenson},
  {Weintraub}, {Fuhrmann}, {Angelakis}, {Zensus}, {Healey}, {Romani}, {Shaw},
  {Grainge}, {Birkinshaw}, {Lancaster}, {Worrall}, {Taylor}, {Cotter}, \&
  {Bustos}}]{ovro}
{Richards}, J.~L., {Max-Moerbeck}, W., {Pavlidou}, V., {et~al.} 2011, \apjs,
  194, 29

\bibitem[{{Saha} {et~al.}(2013){Saha}, {Chitnis}, {Vishwanath}, {Kale},
  {Shukla}, {Acharya}, {Anupama}, {Bhattacharjee}, {Britto}, {Prabhu}, \&
  {Singh}}]{labHAGAR}
{Saha}, L., {Chitnis}, V.~R., {Vishwanath}, P.~R., {et~al.} 2013, Astroparticle
  Physics, 42, 33

\bibitem[{{Sahayanathan} \& {Godambe}(2012)}]{sunder_3c279}
{Sahayanathan}, S., \& {Godambe}, S. 2012, \mnras, 419, 1660

\bibitem[{{Scaringi} {et~al.}(2012){Scaringi}, {K{\"o}rding}, {Uttley},
  {Knigge}, {Groot}, \& {Still}}]{scaringi}
{Scaringi}, S., {K{\"o}rding}, E., {Uttley}, P., {et~al.} 2012, \mnras, 421,
  2854

\bibitem[{{Schlegel} {et~al.}(1998){Schlegel}, {Finkbeiner}, \&
  {Davis}}]{schelgel}
{Schlegel}, D.~J., {Finkbeiner}, D.~P., \& {Davis}, M. 1998, \apj, 500, 525

\bibitem[{{Shukla} {et~al.}(2012){Shukla}, {Chitnis}, {Vishwanath}, {Acharya},
  {Anupama}, {Bhattacharjee}, {Britto}, {Prabhu}, {Saha}, \& {Singh}}]{Amit421}
{Shukla}, A., {Chitnis}, V.~R., {Vishwanath}, P.~R., {et~al.} 2012, \aap, 541,
  A140

\bibitem[{{Sikora} {et~al.}(1994){Sikora}, {Begelman}, \&
  {Rees}}]{leptonic_sikora}
{Sikora}, M., {Begelman}, M.~C., \& {Rees}, M.~J. 1994, \apj, 421, 153

\bibitem[{{Singh} {et~al.}(2014){Singh}, {Tandon}, {Agrawal}, {Antia},
  {Manchanda}, {Yadav}, {Seetha}, {Ramadevi}, {Rao}, {Bhattacharya}, {Paul},
  {Sreekumar}, {Bhattacharyya}, {Stewart}, {Hutchings}, {Annapurni}, {Ghosh},
  {Murthy}, {Pati}, {Rao}, {Stalin}, {Girish}, {Sankarasubramanian},
  {Vadawale}, {Bhalerao}, {Dewangan}, {Dedhia}, {Hingar}, {Katoch}, {Kothare},
  {Mirza}, {Mukerjee}, {Shah}, {Shah}, {Mohan}, {Sangal}, {Nagabhusana},
  {Sriram}, {Malkar}, {Sreekumar}, {Abbey}, {Hansford}, {Beardmore}, {Sharma},
  {Murthy}, {Kulkarni}, {Meena}, {Babu}, \& {Postma}}]{astrosat}
{Singh}, K.~P., {Tandon}, S.~N., {Agrawal}, P.~C., {et~al.} 2014, in Society of
  Photo-Optical Instrumentation Engineers (SPIE) Conference Series, Vol. 9144,
  Society of Photo-Optical Instrumentation Engineers (SPIE) Conference Series,
  1

\bibitem[{{Sinha} {et~al.}(2016){Sinha}, {Shukla}, {Saha}, {Acharya},
  {Anupama}, {Bhattacharjee}, {Britto}, {Chitnis}, {Prabhu}, {Singh}, \&
  {Vishwanath}}]{421_longterm}
{Sinha}, A., {Shukla}, A., {Saha}, L., {et~al.} 2016, \aap, 591, A83

\bibitem[{{Sironi} {et~al.}(2015){Sironi}, {Keshet}, \& {Lemoine}}]{sironi}
{Sironi}, L., {Keshet}, U., \& {Lemoine}, M. 2015, \ssr, 191, 519

\bibitem[{{Sowards-Emmerd} {et~al.}(2003){Sowards-Emmerd}, {Romani}, \&
  {Michelson}}]{egret_NS}
{Sowards-Emmerd}, D., {Romani}, R.~W., \& {Michelson}, P.~F. 2003, \apj, 590,
  109

\bibitem[{{Tavecchio} {et~al.}(1998){Tavecchio}, {Maraschi}, \&
  {Ghisellini}}]{tavKN}
{Tavecchio}, F., {Maraschi}, L., \& {Ghisellini}, G. 1998, \apj, 509, 608

\bibitem[{{Urry} \& {Padovani}(1995)}]{UrryPadovani}
{Urry}, C.~M., \& {Padovani}, P. 1995, \pasp, 107, 803

\bibitem[{{Uttley} \& {McHardy}(2001)}]{lognorm_xrb}
{Uttley}, P., \& {McHardy}, I.~M. 2001, \mnras, 323, L26

\bibitem[{{Vaughan} {et~al.}(2003){Vaughan}, {Edelson}, {Warwick}, \&
  {Uttley}}]{Vaughan}
{Vaughan}, S., {Edelson}, R., {Warwick}, R.~S., \& {Uttley}, P. 2003, \mnras,
  345, 1271

\bibitem[{{Zirakashvili} \& {Aharonian}(2007)}]{zirak}
{Zirakashvili}, V.~N., \& {Aharonian}, F. 2007, \aap, 465, 695

\end{thebibliography}
\bibliographystyle{apj}


%
%
%

\begin{table*}
	\raggedleft
	\scriptsize
	\begin{tabular}{c c c c c c c c c c c }
		\hline
		\hline
		state & Start date & End date & XRT obs id & XRT exp & \multicolumn{3}{c}{XRT spectral parameters} & \multicolumn{2}{c}{LAT spectral parameters} \\
					   &  ISO & ISO	&	   & time (ks) & $\alpha$ & $\beta$ & $F_{2-10\ keV}$ & Index & $F_{0.2-300\ GeV}$\\
		\hline

		\vspace{-0.2cm}
		\multirow{4}{*}{s1} & \multirow{4}{*}{2012-03-19} & \multirow{4}{*}{2012-04-03} & 00035012020 & \multirow{4}{*}{6.3}  & \multirow{4}{*}{$2.27\pm 0.19$} & \multirow{4}{*}{$0.17\pm 0.10$} & \multirow{4}{*}{$1.6 \pm 0.3$} & \multirow{4}{*}{$1.81 \pm 0.16$} & \multirow{4}{*}{$4.3 \pm 0.9$}\\ 
		\vspace{-0.2cm}
				        &							&							& 00035012021 \\
		\vspace{-0.2cm}
							  &							  &				&  00035012022 \\
		\vspace{-0.2cm}
					&							 &					&		  00035012023 \\

		\multirow{5}{*}{s2} & \multirow{5}{*}{2012-12-19} & \multirow{5}{*}{2013-01-18} & 00035012024 & \multirow{5}{*}{5.1}  & \multirow{5}{*}{$2.37\pm 0.11$} & \multirow{5}{*}{$0.61\pm 0.22$}  & \multirow{5}{*}{$0.82 \pm 0.07$} & \multirow{5}{*}{$1.74 \pm 0.13$} & \multirow{5}{*}{$2.8 \pm 0.6$}\\ 
		\vspace{-0.2cm}
				        &							&							& 00035012025 \\
		\vspace{-0.2cm}
							  &							  &				&  00035012026 \\
		\vspace{-0.2cm}
							  &							 &					&		  00035012027 \\
		\vspace{-0.2cm}
				        &							&							& 00035012028 \\
		\vspace{-0.2cm}
							  &							  &				&  00035012029 \\
	
		\multirow{6}{*}{s3} & \multirow{6}{*}{2014-02-04} & \multirow{6}{*}{2014-03-04} & 00035012030 & \multirow{6}{*}{7.3} & \multirow{6}{*}{$1.94\pm 0.04$} & \multirow{6}{*}{$0.16\pm 0.0.03$}  & \multirow{6}{*}{$4.11 \pm 0.12$}&\multirow{6}{*}{$1.77 \pm 0.22$} &\multirow{6}{*}{$9.8 \pm 0.3$}  \\ 
		\vspace{-0.2cm}
				        &							&							& 00035012031 \\
		\vspace{-0.2cm}
							  &							  &				&  00035012032 \\
		\vspace{-0.2cm}
							  &							 &					&		  00035012033 \\
		\vspace{-0.2cm}
				        &							&							& 00035012035 \\
							  &							  &				&  00035012036 \\

		\hline
		\hline
	\end{tabular}
	\caption{Observation details (XRT observation ids and the total exposure time) and spectral parameters in the X-ray and GeV bands for the different states for which the SED has been extracted. The X-ray $2-10$ keV flux is quoted in $10^{-11}$ ergs/cm$^2$/s, and the Fermi-LAT $0.2-300$ GeV flux in $10^{-8}$ ph/cm$^2$/s.}
	\label{tab:states}
\end{table*}

\begin{table*}
	\centering
	\scriptsize
	\begin{tabular}{ccccccccc}
		\hline
		\hline
		\multicolumn{9}{c} {Broken Power Law (BPL)}\\
		{State} & \multicolumn{2}{c} {Particle index} & Magnetic Field & Doppler Factor & Break Energy & Particle energy density & Luminosity & $\chi^2$/dof\\
						  &    p   &   q & $B$ ($G$)   & 	$\delta$ &  $\gamma_{b}$ 	& $U_e$ (ergs/cc) & $L$ (ergs/cc) & \\
		\hline 

		s1 & 2.35 &  4.20  & 0.82 &     10.0  &    7.7e4   & 5.1e-2 &  3.2e46 & 10.3\\
		s2 & 2.20  & 4.60  & 0.78  &     10.2 & 		8.1e4 &  3.8e-2 & 2.6e46 & 9.4\\
		s3 & 2.26  & 4.30 & 0.73  &     9.8   &     1.9e5  &  7.4e-2 &  7.7e46 & 12.1\\

		\hline
		\hline
		\multicolumn{9}{c} {Smooth Broken Power Law (SBPL)}\\
		&    p   &   q & $B$ ($G$)   & 	$\delta$ &  	$\gamma_{b}$ & $U_e $ (ergs/cc)  & $L$ (ergs/cc) & \\
		\hline 
		s1 & 2.35   &     4.22   &     0.83   &     10.0   &    9.5e4  &  7.8e-2  &  3.7e46 & 1.3 \\
		s2 & 2.20   &     4.60   &    0.78   &      10.2   &    7.7e4   &  4.7e-2 &  2.3e46 & 1.1 \\
		s3 & 2.22   &     4.20   &     0.73  &     9.8   &     1.7e5  &  8.6e-2 &  7.8e46 & 1.2 \\
		\hline
		\hline
		\multicolumn{9}{c} {Cutoff Power Law (CPL)}\\
		&    \multicolumn{2}{c}p    & $B$ ($G$)   & 	$\delta$ &  $\gamma_{max}$	& $U_e$ (ergs/cc) & $L$ (ergs/cc) &  \\
		\hline
		s1    &  \multicolumn{2}{c}{2.30}  & 0.78  	& 10.9    &  1.1e5 	&		8.2e-2 & 3.4e46 &  1.2\\
		s2    &  \multicolumn{2}{c}{2.02} & 0.76     & 10.7   &  7.0e4 	&	    4.6e-2 &  2.1e46   &  1.4 \\ 
		s3    &	 \multicolumn{2}{c}{2.10} & 0.74    & 10.3  &  1.6e5 &       8.6e-2 &  6.6e46 &  1.3 \\
		\hline
		\hline

	\end{tabular}
	\caption{Models parameters, the total bolometric luminosity ($L$) and the computed reduced-$\chi^2$ for the different particle distributions during the three epochs. While the BPL cannot reproduce the observed spectrum satisfactorily, the SBPL and the CPL can. }  
	\label{partable}
\end{table*}


\end{document}